\documentclass[aps,superscriptaddress,twocolumn]{revtex4-2}
\usepackage[T1]{fontenc}
\usepackage{amsmath, amsfonts, mathtools}
\usepackage{textcomp}
\usepackage[utf8]{inputenc}
\usepackage{graphicx}
\usepackage{float}
\usepackage[caption=false]{subfig}
\usepackage{geometry}
\usepackage{comment}
\usepackage{babel}
\usepackage[dvipsnames]{xcolor}
\usepackage{braket}
\usepackage[normalem]{ulem}

\usepackage{mleftright, soul}

\definecolor{adon}{RGB}{0,155,185}
\newcommand{\adon}[1]{\textcolor{adon}{#1}}
\definecolor{lkc}{RGB}{0,155,0}
\newcommand{\lkc}[1]{\textcolor{lkc}{#1}}

\begin{document}
\title{Minimal-Energy Optimal Control of Tunable Two-Qubit Gates in Superconducting Platforms Using Continuous Dynamical Decoupling}
\author{Adonai Hilário da Silva}
\affiliation{Sao Carlos Institute of Physics, University of Sao Paulo, IFSC – USP,
13566-590, Sao Carlos, SP, Brazil}

\author{Octávio da Motta}
\affiliation{Sao Carlos Institute of Physics, University of Sao Paulo, IFSC – USP,
13566-590, Sao Carlos, SP, Brazil}

% \author{Gabriel Nogueira Audi Monteiro}
% \affiliation{Sao Carlos Institute of Physics, University of Sao Paulo, IFSC – USP, 13566-590, Sao Carlos, SP, Brazil}

% \author{Nicolas André da Costa Morazotti}
% \affiliation{Sao Carlos Institute of Physics, University of Sao Paulo, IFSC – USP, 13566-590, Sao Carlos, SP, Brazil}
% \affiliation{Departamento de Física, Universidade Federal de São Carlos (UFSCar), São Carlos, SP 13565-905, Brazil}

\author{Leonardo K. Castelano}
\email{lkcastelano@ufscar.br}
\affiliation{Departamento de Física, Universidade Federal de São Carlos (UFSCar),
São Carlos, SP 13565-905, Brazil}

\author{Reginaldo de Jesus Napolitano}
\affiliation{Sao Carlos Institute of Physics, University of Sao Paulo, IFSC – USP,
13566-590, Sao Carlos, SP, Brazil}

\begin{abstract}
We present a unified scheme for generating high-fidelity entangling gates in superconducting platforms by continuous dynamical decoupling (CDD) combined with variational minimal-energy optimal control. During the CDD stage, we suppress residual couplings, calibration drifting, and quasistatic noise, resulting in a stable effective Hamiltonian that preserves the designed ZZ interaction intended for producing tunable couplers. In this stable $\mathrm{SU}\left(4\right)$ manifold, we calculate smooth low-energy single-quibt control functions using a variational geodesic optimization process that directly minimizes gate infidelity. We illustrate the methodology by applying it to CZ, CX, and generic engangling gates, achieving virtually unit fidelity and robustness under restricted single-qubit action, with experimentally realistic control fields. These results establish CDD-enhanced variational geometric optimal control as a practical and noise-resilient scheme for designing superconducting entangling gates.
\end{abstract}
\maketitle

\section{Introduction}

Recent developments in quantum processing capabilities promise advantages
over classical computers when quantum gates transcend the current
noisy intermediate-scale quantum (NISQ) operation regime~\cite{Lau2022,10460214,Preskill2018quantumcomputingin}.
One- and two-qubit gates in the NISQ era are plagued by various noise
sources. Today, entangling two-qubit gates, such as the controlled-Z
(CZ) gate, are considered stepping stones to scalable quantum computing~\cite{RevModPhys.87.307,PhysRevA.86.032324}.
Research groups such as IBM and Google have built CZ gates of impressive
high fidelities using tunable couplers in superconducting transmon
qubits~\cite{PhysRevX.11.021058,PhysRevX.14.041050,annurev-conmatphys-031119-050605,PhysRevResearch.2.033447,PhysRevLett.125.120504}.
To overcome coherent control errors, unwanted ZZ crosstalk between
qubits, and leakage to higher transmon levels, CZ-gate implementations
employ flux-pulse shaping, echo sequences, and derivative removal
by adiabatic gate (DRAG)~\cite{PhysRevLett.103.110501,PhysRevA.93.060302,Jurcevic_2021,PhysRevA.90.022307,PhysRevA.82.040305,8lxc-lvv1,PhysRevA.83.012308}.
Despite the high fidelities achieved through these strategies,
multiple layers of correction are involved, such as pulse shaping, dynamical
decoupling echoes, etc., and, as device parameters drift, frequent
recalibration is required~\cite{PhysRevLett.123.190502,10.1145/3695053.3731036}.

However, several two-qubit vulnerabilities remain that need to be
overcome. Low-frequency flux fluctuations, calibration drift, and
other slow coherent noise sources can accumulate phase errors that
resist the usual echo pulses~\cite{10.1145/3695053.3731036,PhysRevLett.130.220602,PhysRevApplied.12.054015}. Residual, always-on ZZ couplings also
destroy gate fidelity and require continuous suppression~\cite{8lxc-lvv1,PhysRevLett.125.200504,PhysRevX.11.021058}. Moreover,
slight pulse miscalibration causes leakage to non-computational states,
such as the second excited state in a transmon, which limits fast
entangling-gate fidelities~\cite{PhysRevLett.103.110501,PhysRevLett.123.120502,PhysRevLett.126.220502}. Addressing
these issues individually resulted in a complex and unsystematic control
protocol. A crucial open question is whether it is possible to devise
a unified and systematic control scheme that implements a high-fidelity
entangling gate while suppressing slow noise and spurious couplings.

Here, we introduce a strategy to implement robust two-qubit entangling gates under realistic
noise in two stages. In the first stage, we use dynamical decoupling (a relevant and thorough review focusing on superconducting platforms can be found in~\cite{lidar}). Specifically, for our present purposes, we use continuous dynamical decoupling
(CDD)~\cite{Cai_2012} to preemptively eliminate low-frequency noise and spurious static
terms in the Hamiltonian. Instead of performing discrete refocusing
pulses, this stage continuously applies driving control fields that
effectively average out undesired perturbations during gate execution~\cite{PhysRevResearch.6.013217,PhysRevLett.121.130501}.
In the second stage, we add optimized control fields to drive the effective
two-qubit gate. This set of control fields
steers the evolution along a geodesic trajectory in $\mathrm{SU}(4),$ generated by a control Hamiltonian projected from an auxiliary co-state matrix $\Lambda(0).$ This single parametric matrix determines the entire time dependence of the control fields through a nonlinear Schrödinger equation. To optimize gate fidelity, we introduce a second Lagrange multiplier matrix, $\Gamma(t) $, propagated backward in time from a terminal condition derived from the fidelity gradient. This second-variation principle yields an efficient boundary value optimization over $\Lambda(0) $ that does not require neural networks, pulse shaping, or time-dependent parameterization.

We obtain high-fidelity two-qubit entangling gates in our simulations using our unified
control scheme. The simulations show robustness against coherent noise
and crosstalk. Unlike multi-layered approaches that separately address different error sources, our methodology achieves robustness and control effectiveness in a unified process through a single optimization over $\Lambda(0).$ For simplicity,
we model the superconducting transmons as ideal two-state qubits, ignoring leakage.
However, the variational framework can be naturally extended to higher-dimensional control spaces. In a future investigation on how to suppress leakage during the operation of fast gates,
we will extend the present $\mathrm{SU}(4)$ model to
$\mathrm{SU}(9)$.

\section{Physical Model and Effective Hamiltonian}

\subsection{Two-qubit Hamiltonian}
 Let us start with the representation of the laboratory, the lab frame.
From this point of view, we assume a free evolution for the two qubits:
\begin{eqnarray}
H_{0} & = & \omega_{z}\sigma_{z}\otimes I+\omega_{z}I\otimes\sigma_{z},\label{H0}
\end{eqnarray}
where we take the typical value of $5\text{ GHz}$ for $\omega_{z}/2\pi$~\cite{Place2021,Chen2023} and henceforth we use units of angular frequency instead of energy. For
the sake of concreteness, we assume that a ZZ coupling is tuned
intentionally at its maximum fixed value, $J_{z,z}/2\pi=82.5\text{ MHz}$, as in~\cite{PhysRevX.14.041050},
where a double-transmon coupler is investigated in detail, reaching
remarkable fidelities.

Due to the large separation between the qubit frequency $\omega_{z}$ and the interaction strength $J_{z,z},$ it is standard in superconducting qubit models to transform to the so-called rotating frame~\cite{RevModPhys.93.025005}, defined by the qubit eigenfrequencies. This eliminates fast oscillations and enables an effective description in which ideal qubits are gapless, only interactions and control fields contribute to the dynamics. The resulting frame isolates the relevant terms of entanglement and control, greatly simplifying gate design and analysis. We accomplish this change of point of view by performing
a unitary transformation using the unitary operator
\begin{eqnarray}
U_{0}(t) & = & \exp\mleft(-iH_{0}t\mright),\label{U0}
\end{eqnarray}
where $t$ is the time measured from the ZZ coupling
turning on at $t=0$, and $H_{0}$ is given by Eq.~(\ref{H0}).

Let us consider, from the rotating point of view, a general native Hamiltonian in the absence of any control fields given by

\begin{eqnarray}
H_{N} & = & \sum_{\mu=0}^{3}\sum_{\nu=0}^{3}J_{\mu,\nu}\sigma_{\mu}\otimes\sigma_{\nu},\label{nativeH}
\end{eqnarray}
where $\sigma_{0}=I,$ $I$ is the identity matrix, and $\left(\sigma_{1},\sigma_{2},\sigma_{3}\right)=\left(\sigma_{x},\sigma_{y},\sigma_{z}\right),$ are the Pauli matrices. Because we restrict the analysis to $\mathrm{SU}(4),$
we assume $J_{0,0}=0,$ so that the Hamiltonian is traceless. We take
all $J_{\mu,\nu}$ as real constants for definiteness. In the rotating
frame, the term $J_{3,3}\sigma_{3}\otimes\sigma_{3}$ is invariant
and we have the same coupling constant $J_{3,3}=J_{z,z}$ as tuned at $t=0$
in the lab frame.

We emphasize that this native Hamiltonian, Eq.~(\ref{nativeH}), represents the internal deterministic evolution generated by the quantum platform, without any external control fields or environmental couplings. It includes engineered two-qubit interactions (including tunable couplings) and intrinsic local terms that arise from device fabrication and coupling schemes. It does not yet account for stochastic effects such as
dephasing, relaxation, or other forms of environmental noise. These effects will be addressed by control strategies designed to reduce their impact during gate operation. Moreover, we emphasize that Eq. (\ref{nativeH}) is expressed from the rotating-frame point of view and that the rotating-wave approximation has already been applied to any fast-varying terms.

The general form of Eq.~(\ref{nativeH}) encompasses all bilinear
two-qubit interactions as well as local drift terms. The terms $\left(\mu,\nu\right)=\left(\mu\neq0,0\right)$
or $\left(\mu,\nu\right)=\left(0,\nu\neq0\right)$ describe the single-qubit evolutions of the first and second qubits, respectively. Terms with both indices different from zero describe two-qubit interactions. Equation (\ref{nativeH}) thus captures a wide range of physically
relevant interactions, such as engineered and residual ZZ, capacitive XX or YY couplings, etc. The assumed relative magnitudes of the $J_{\mu,\nu}$ coupling constants will be discussed shortly.

We emphasize the important fact that this formulation is in full alignment with the $15\text{-dimensional}$ vector space generated by the Lie algebra $\mathfrak{su}(4),$ in a way that it is naturally compatible with geodesic control strategies~\cite{PhysRevA.110.042601,dasilva2025montecarloapproachfinding}.
Equation (\ref{nativeH}), which does not include control fields,
induces a baseline dynamics, which will be modified by the introduction of continuous control fields to suppress noise and implement high-fidelity two-qubit entangling gates. The flexibility of this representation allows us to encode
a variety of realistic native Hamiltonians found in superconducting
and other quantum computing platforms, while keeping the formalism
fully compatible with the differential geometric tools we employ. Except for the intended coupling $J_{3,3}$, 
the remaining $J_{\mu,\nu}$ couplings will be taken as residual, due to spurious couplings resulting from eventual fabrication imperfections and peculiarities inherent to the platform and tuning scheme. Thus, we assume a maximum magnitude
of $10\text{ MHz}$ if $\mu\neq\nu$ and $50\text{ MHz}$
if $\mu=\nu=1,2$ for these remaining $J_{\mu,\nu}/2\pi$ couplings, consistent with typical residual couplings observed in superconducting circuits ~\cite{PhysRevX.11.021058,PhysRevX.14.041050,annurev-conmatphys-031119-050605,PhysRevA.93.060302,PhysRevLett.123.120502}.
To fix the present discussion, we choose a $4\times4$ matrix $\mathbf{J}$
with entries $J_{\mu,\nu}$ given by
\begin{eqnarray*}
\frac{\mathbf{J}}{2\pi} & = & \left(\begin{array}{cccc}
0 & 1.1 & 1.2 & 5.7\\
4.7 & 11.8 & 9.4 & 2.1\\
4.8 & 9.8 & 31.6 & 3.4\\
0.8 & 9.0 & 0.2 & 82.5
\end{array}\right)\text{ MHz}.
\end{eqnarray*}

Current quantum computing systems involve many qubits and gates. Each two-qubit system, during the action of a quantum gate, gets perturbed by neighboring qubits being actuated by parallel gate operations driven by external control fields. These additional unwanted couplings are commonly referred to as crosstalk. These perturbations are generally weaker than the gate couplings of the system in focus, but can accumulate coherent phase errors or drive small population leakage, resulting in poorer gate fidelity~\cite{annurev-conmatphys-031119-050605,Jurcevic_2021}. Within the subspace of the two qubits in focus, crosstalk with qubits from other subspaces can be approximated, during the gate operation within the focused subspace, as additional fixed $J_{\mu,\nu}\text{-like}$ couplings. Hence, we assume that these additional couplings are already included by our choice of maximum values for the couplings $J_{\mu,\nu}$, except $J_{3,3},$ since the crosstalk coupling strengths are usually weaker than $1\text{ MHz},$ as suggested by experimental tendencies~\cite{Jurcevic_2021} and the known characteristics of residual coupling mechanisms~\cite{annurev-conmatphys-031119-050605}.

Although $H_{N},$ Eq.~(\ref{nativeH}), can be chosen within a gate duration $\tau$, for a fixed $\mathbf{J}$ matrix as above, in reality the $J_{\mu,\nu},$ except for $J_{3,3},$ are not constant. Fluctuations from crosstalk, calibration drift, or activity-dependent interactions can cause $\mathbf{J}$ to vary between gate executions or when two similar gates act at other circuit sites. Therefore, control fields optimized for a fixed choice of $\mathbf{J}$ will have a lower fidelity for another realization of $\mathbf{J}.$ Our CDD method solves this difficulty by dynamically suppressing all Hamiltonian components other than the component $J_{3,3},$ as discussed in Sec.~\ref {sec:Stage-1:-Continuous}, effectively filtering out the
random spurious terms. As we shall see, the remaining averaged Hamiltonian (see Eq.~(\ref{avgHN}), below) becomes more stable and predictable, enabling control fields to remain effective over repeated gate use and circuit execution.

To further clarify this point, the ZZ coupling term $J_{3,3}$ in the native Hamiltonian, Eq.~(\ref{nativeH}), is the only interaction intentionally engineered for entanglement and is typically well-characterized and tunable for superconducting platforms, as exemplified in~\cite{PhysRevX.14.041050}. Although small drifts may occur during long timescales, $J_{3,3}$ remains effectively constant during gate execution and each time the gate is executed. We therefore treat $J_{3,3}$ as fixed during optimization and include it as the sole retained interaction in the effective drift Hamiltonian, after dynamical decoupling, as explained in Sec.~\ref{sec:Stage-1:-Continuous}.

\subsection{Environmental Noise Model}

The environment of a qubit system is also expected to introduce noise in the desired gate operation, causing dissipation and decoherence mainly through single-qubit couplings to external fields, typically modeled as thermal baths of bosons~\cite{Breuer2007-ti}. Thus, to include these environmental perturbations, we define the global Hamiltonian in the rotating frame as
\begin{eqnarray}
H_\mathrm{gl} & = & H_{N}\otimes I_{E}+I\otimes I\otimes H_{E}+H_\mathrm{int},\label{Hgl}
\end{eqnarray}
where $I_{E}$ is the identity operator acting on the states belonging
to the environmental Hilbert space, $H_{N}$ is given by Eq.~(\ref{nativeH}),
and $H_{E}$ is the environmental Hamiltonian. The term $H_\mathrm{int}$ is the interaction between the environment and the two qubits. In the laboratory representation, the environmental interaction is assumed to be given by
\begin{eqnarray}
H_\mathrm{int}^{(\mathrm{lab})} & = & \sum_{k=1}^{3}\left(\sigma_{k}\otimes I\otimes E_{k}^{(1)}+I\otimes\sigma_{k}\otimes E_{k}^{(2)}\right),\label{Hlab}
\end{eqnarray}
where $E_{k}^{(1)}$ and $E_{k}^{(2)}$ are
Hermitian environmental operators. Since we assume identical qubits, their environments, although independent, are also assumed to be identical. Given the assumption of independence, we use superscripts in $E_{k}^{(1)}$ and $E_{k}^{(2)}$, but these operators are not the same for different subscripts in general. Because we transform $H_\mathrm{int}^{(\mathrm{lab})}$ using Eqs.~(\ref{H0}) and (\ref{U0}), the $\sigma_{1}=\sigma_{x}$ and $\sigma_{2}=\sigma_{y}$ operators appearing in Eq.~(\ref{Hlab})
will rotate with frequency $\omega_{z}$ in the rotating frame, and we assume that the magnitudes of the noise operators $E_{k}^{\left(j\right)}$ are much smaller than $\omega_{z},$ guaranteeing the validity of the rotating-wave approximation. Hence, from the rotating point of view, we only keep the resulting interaction Hamiltonian:
\begin{eqnarray}
H_\mathrm{int} & = & \sigma_{3}\otimes I\otimes E_{3}^{(1)}+I\otimes\sigma_{3}\otimes E_{3}^{(2)}.\label{Hint}
\end{eqnarray}
We take the standard point of view of the rotating-frame representation~\cite{RevModPhys.93.025005} henceforth,
unless otherwise stated.

\begin{comment}
Equation (\ref{master}) in Appendix \ref{sec:Derivation-of-the}, after proper algebraic manipulation, assumes a Redfield form given by~\cite{Breuer2007-ti} 
\begin{eqnarray}
\frac{d}{\mathrm{d}t}\rho_{SI}(t) & \approx & -\sum_{s=1}^{2}\left[S_{3}^{\left(s\right)}(t),R_{s}^{\dagger}(t)\rho_{IS}(t)\right]\nonumber \\
 &  & +\sum_{s=1}^{2}\left[S_{3}^{\left(s\right)}(t),\rho_{IS}(t)R_{s}(t)\right],\label{Redfield}
\end{eqnarray}
with
\begin{eqnarray*}
R_{s}(t) & = & \int_{0}^{t}\mathrm{d}t_{1}C(t-t_{1})S_{3}^{\left(s\right)}\left(t_{1}\right),
\end{eqnarray*}
where the correlation function is defined by
\begin{eqnarray*}
C(t-t_{1}) & = & \mathrm{Tr}_{E}\left[E_{3}^{(1)}(t)\rho_{E}E_{3}^{(1)}\left(t_{1}\right)\right]\\
 & = & \mathrm{Tr}_{E}\left[E_{3}^{(2)}(t)\rho_{E}E_{3}^{(2)}\left(t_{1}\right)\right],
\end{eqnarray*}
given that $E_{3}^{(1)}$ and $E_{3}^{(2)}$
are identical fields for qubits $1$ and $2,$ though independent
(uncorrelated). Here, the qubit operators are defined as
\begin{eqnarray}
S_{3}^{(1)}(t) & = & U_{N}^{\dagger}(t)\sigma_{3}\otimes IU_{N}(t)\label{S1}
\end{eqnarray}
and
\begin{eqnarray}
S_{3}^{(2)}(t) & = & U_{N}^{\dagger}(t)I\otimes\sigma_{3}U_{N}(t),\label{S2}
\end{eqnarray}
with $U_{N}(t)$ defined by
\begin{eqnarray}
U_{N}(t) & = & \exp\left(-iH_{N}t\right).\label{UN}
\end{eqnarray}
\end{comment}

In superconducting qubit platforms, gate durations typically fall
between $40$ and $70\text{ns}$~\cite{PhysRevX.11.021058,PhysRevX.14.041050,jin2024fasttunablehighfidelity}, much shorter than the usual environmental correlation times. We make a quasistatic noise approximation. Therefore, we assume that the noise correlation function $C\mleft(t-t_{1}\mright)$ remains effectively constant, that is, $C\mleft(t-t_{1}\mright)\approx C(0),$ during gate operation.

The constant $C(0)$ sets the magnitude of the environmental noise, and is related to the free-induction decay time $T_{2}^{\ast},$ which captures the decoherence due to both true dephasing and slow fluctuations in the system. In typical experiments with transmon qubits, $T_{2}^{\ast},$ as measured by Ramsey interferometry, without echo correction, can be found between $10$ and $50\text{ \ensuremath{\mu}s}$, and some devices fall out of this range~\cite{PhysRevLett.107.240501,PhysRevB.86.100506,PhysRevLett.111.080502,doi:10.1126/sciadv.aao3603}. Here, we choose $T_{2}^{\ast}=20\text{ \ensuremath{\mu}s}$ as a conservative
value. Our approach models decoherence by a second-order Redfield
equation under the assumption of constant environmental correlation, giving a Gaussian decay of the coherence that behaves as $\rho_{01}(t)=\rho_{01}(0)\exp\mleft(-\gamma t^{2}\mright)$
for a single qubit, without any perturbations and in the rotating frame. We adjust this to the observed free induction decay times in superconducting platforms by setting $\gamma=1/\left(T_{2}^{\ast}\right)^{2}=2.5\times10^{9}\text{ s}^{-2}.$

A realistic decoherence rate, such as $\gamma$, induces only a minor impact on the average fidelity,
% as explained in Sec.~\ref{subsec:Effective-Drift-Hamiltonian} (see Eq.~(\ref{Fbar})),
if we choose $C(0)=\gamma$
%in Eq.~(\ref{Redfield})
in a standard Redfield equation and solve it for our gate time choice of $\tau=40\text{ ns},$
as is the case for realistic superconducting gate durations (e.g. $40-70\text{ ns}$).
\begin{comment}
However, even such a weak noise accumulates during
the full execution of large quantum circuits, which might require
thousands of quantum gates of similar duration, including the idle
time intervals between adjacent gate actions. To model this cumulative
vulnerability within our single-qubit analysis, we choose the noise
strength in Eq.~(\ref{Redfield}), $C(0),$ by a factor
of $10^{3}-10^{4}$ times $\gamma.$ Our intention with this approach
is to compress the impact of long-term decoherence into the gate duration
window. We are then able to assess our scheme suppresses errors that
would otherwise accumulate throughout the many idle or gate cycles
involved in a realistic quantum circuit application. Our procedure
is not arbitrary, but rather a device to incorporate future demands
of systemic coherence withing the control design of today.

As we shall show,
\end{comment}
Therefore, we neglect such a small environmental noise during gate operation, because, as it will become apparent from the discussion in Sec.~\ref{sec:Stage-1:-Continuous}, the same continuous dynamical decoupling (CDD) field
used to suppress spurious $\mathbf{J}$ components, given a suitable
choice of frequency $\omega,$ turns out to be also effective to suppress such a weak low-frequency environmental noise. An always-on CDD field scheme will
also inherently suppress the cumulative decoherence that would otherwise
build up over idle intervals and long quantum-circuit depths. The CDD control we propose has a twofold role:
mitigating error mechanisms arising both from unitary imperfections
and non-unitary environmental decoherence.

\section{Continuous Dynamical Decoupling (CDD)}\label{sec:Stage-1:-Continuous}

\subsection{CDD Transformation and Effective Hamiltonian}\label{subsec:CDD-Transformation-and}

We adopt a version of CDD inspired by Ref.~\cite{PhysRevA.110.032607},
defining a unitary transformation as
\begin{eqnarray}
U_\mathrm{CDD}(t) & = & U_{1}(t)\otimes U_{2}(t),\label{UCDD}
\end{eqnarray}
where
\begin{eqnarray*}
U_{1}(t) & = & \exp\mleft(-i\omega t\sigma_{z}\mright)\exp\mleft(-3i\omega t\sigma_{x}\mright)
\end{eqnarray*}
and
\begin{eqnarray*}
U_{2}(t) & = & \exp\mleft(-8i\omega t\sigma_{z}\mright)\exp\mleft(-3i\omega t\sigma_{x}\mright).
\end{eqnarray*}
We notice that this is a modified version of the one used in Ref.~\cite{PhysRevA.110.032607},
because, by using the frequency $3\omega$ rather than $2\omega$
in the second exponential factor defining $U_{1,2}(t)$ and $8\omega$ in the first exponential in $U_{2}(t),$ we obtain an effective Hamiltonian that preserves hardware-compatible
entangling interactions, as we will explain next. These modifications ensure that the following result holds:
\begin{eqnarray}\label{Hd}
H_{d} & = & \frac{\omega}{2\pi}\int_{0}^{2\pi/\omega}\mathrm{d}t\,\widetilde{H}_\mathrm{gl}(t) \nonumber \\
& = & \frac{J_{3,3}}{2}\sigma_{y}\otimes\sigma_{y}\otimes I_{E}+\frac{J_{3,3}}{2}\sigma_{z}\otimes\sigma_{z}\otimes I_{E}, \nonumber \\
&   & +\, I \otimes I \otimes H_E \label{avgHN}
\end{eqnarray}
where $\widetilde{H}_\mathrm{gl}(t)$ is the global Hamiltonian of
Eq.~(\ref{Hgl}) unitarily transformed, i.e.,
\begin{eqnarray}
\widetilde{H}_\mathrm{gl}(t) & = & \left[U_\mathrm{CDD}^{\dagger}(t)\otimes I_{E}\right]H_\mathrm{gl}\left[U_\mathrm{CDD}(t)\otimes I_{E}\right].\label{CDDtransform}
\end{eqnarray}
Indeed, we explicitly see in Eq.~(\ref{avgHN}) the retention of the
relevant capacitive, $\sigma_{y}\otimes\sigma_{y},$
and dispersive, $\sigma_{z}\otimes\sigma_{z},$ coupling terms. We can also see that this Hamiltonian describes the system and the environment evolving without any interaction, so we can effectively ignore the term $I \otimes I \otimes H_E$. Moreover, as we emphasized at the end of the previous section, only the stable $J_{3,3}$ component remains present, after our deliberate choice of the CDD scheme, Eq.~(\ref{UCDD}). This
is noteworthy since the remaining terms in Eq.~(\ref{avgHN}) are of the kind that naturally occurs in superconducting qubit
platforms, implemented as exchange interactions and residual static
shifts, respectively. For high-fidelity two-qubit gates, the resulting effective Hamiltonian is characteristically separated from the low-frequency
noise and undesirable single-qubit rotations, increasing experimental
maneuverability and robustness without the introduction of additional
complicated calibration schemes.

\subsection{Fidelity Benchmarking and the $\omega$ Selection}\label{subsec:Fidelity-Benchmarking-and}

To determine the CDD frequency that implements the effective transformation of the system described by $H_\mathrm{gl}$ of Eq.~(\ref{Hgl}) to the one described by the drift Hamiltonian of Eq.~(\ref{avgHN}), we proceed as follows.

We start by solving the
ideal evolution by the drift Hamiltonian, Eq.~(\ref{avgHN}), without CDD and comparing it against the noisy case, Eq.~(\ref{Hgl}), in the presence of the CDD. Namely, we solve
\begin{eqnarray*}
i\frac{\mathrm{d}U_\mathrm{id}(t)}{\mathrm{d}t} & = & H_{d}U_\mathrm{id}(t),
\end{eqnarray*}
with $U_\mathrm{id}(0)=I\otimes I,$ and obtain the ideal $U_\mathrm{id}(\tau).$ From now on, we ignore the tensor product with the environmental Hilbert space, as it will soon become evident that the CDD fields, used to implement the averaged dynamics of Eq.~(\ref{avgHN}), will also decouple the qubits from the environmental interaction.
We then calculate the test unitary by solving the following:
\begin{eqnarray}
i\frac{\mathrm{d}U_\mathrm{tst}(t)}{\mathrm{d}t} & = & \left[H_{N}+H_\mathrm{CDD}(t)\right]U_\mathrm{tst}(t),\label{Utst}
\end{eqnarray}
where $H_\mathrm{CDD}(t)$ is given by
\begin{eqnarray}
H_\mathrm{CDD}\left(t\right) & = & H_\mathrm{CDD}^{\left(1\right)}\left(t\right)\otimes I+I\otimes H_\mathrm{CDD}^{\left(2\right)}\left(t\right),\label{HCDD}
\end{eqnarray}
with
\begin{eqnarray}
H_\mathrm{CDD}^{\left(1\right)}\left(t\right) & = & \hbar\omega\sigma_{z}+3\hbar\omega\sigma_{x}\cos\left(2\omega t\right)\nonumber \\
 &  & +3\hbar\omega \sigma_{y}\sin\left(2\omega t\right)
\end{eqnarray}
and
\begin{eqnarray}
H_\mathrm{CDD}^{\left(2\right)}\left(t\right) & = & 8\hbar\omega\sigma_{z}+3\hbar\omega\sigma_{x}\cos\left(16\omega t\right)\nonumber \\
 &  & +3\hbar\omega\sigma_{y}\sin\left(16\omega t\right).
\end{eqnarray}

We compare the two unitary operators obtained as described above using the following fidelity measure:
\begin{eqnarray}\label{fidelity-CDD}
\mathcal{F}_\mathrm{CDD} & = & \frac{1}{16}\left|\mathrm{Tr}\mleft[U_\mathrm{id}^{\dagger}U_\mathrm{tst}(\tau)\mright]\right|^2.
\end{eqnarray}
Using $\omega/2\pi=2.0\text{ GHz},$ we get $\mathcal{F}_\mathrm{CDD}\approx0.99807888.$ For comparison,
if we solve Eq.~(\ref{Utst}) without the CDD fields, we get $\mathcal{F}_\mathrm{CDD}\approx0.62568233.$
If we increase the CDD frequency to $\omega/2\pi=10.0\text{ GHz},$
we obtain $\mathcal{F}_\mathrm{CDD}\approx0.99992314,$ and, for $\omega/2\pi=20.0\text{ GHz},$
the fidelity becomes $\mathcal{F}_\mathrm{CDD}\approx0.99998078.$ Here we
confirm that using a CDD frequency in the microwave range, we can render the
resulting effective Hamiltonian in the form of Eq.~(\ref{avgHN}) with high-enough fidelity.

\section{Geodesic-Informed Optimal two-qubit-entangling Gates}

The connection between quantum computation and geometry was first proposed by Nielsen et al.~\cite{doi:10.1126/science.1121541}, where a quantum circuit is reinterpreted as the problem of finding a geodesic in a curved Riemannian manifold. Later, Wang et al.~\cite{PhysRevLett.114.170501} demonstrated that the quantum brachistochrone problem is equivalent to finding a geodesic in the space of unitary operators under an appropriate Riemannian metric. Contributions from our research group started with the derivation of the geodesic differential equation obtained through the minimization of the energy of the optimal control Hamiltonian~\cite{PhysRevA.110.042601}. Within this formulation, the co-state at the initial time $\Lambda(0)$ is the unknown quantity depending on 6 and 15 parameters for a one and two qubits, respectively. By enforcing that the infidelity of a desired quantum gate at the final time of evolution is minimum, a constraint to determine $\Lambda(0)$ can be obtained. Although this extra condition allows us to find 
$\Lambda(0)$, the numerical treatment for this problem is not trivial. To circumvent this difficulty, first, a neural network was successfully used to find $\Lambda(0)$ for single-qubit gates in the presence of dephasing noise~\cite{PhysRevA.110.042601}. To further improve the efficiency of finding $\Lambda(0)$, we developed a faster method based on the Monte Carlo~\cite{dasilva2025montecarloapproachfinding} approach combined with the minimization of the infidelity, which was applied to one- and two-qubit systems with great performance. In the present paper, we address the numerical challenge by considering a new approach that combines both constraints of minimum energy and minimum infidelity into a single functional. We accomplish this by minimizing the infidelity under the constraint that the dynamics satisfy the non-linear geodesic equation, instead of the Schrödinger equation. Furthermore, we deduce the Euler-Lagrange equations that allow the determination of $\Lambda(0)$ in a self-consistent fashion as explained below.

\subsection{Geometric Control Framework}

Once we have established the drift Hamiltonian, Eq.~(\ref{avgHN}), in the CDD frame, defined by the unitary transformation of Eq.~(\ref{CDDtransform}), we consider another control field that acts only on individual qubits. Thus, we have to optimize a time-dependent Hamiltonian:
\begin{eqnarray*}
H_\mathrm{opt}(t) & = & H_{d}+H_{c}(t),
\end{eqnarray*}
where the control Hamiltonian is defined by
\begin{eqnarray}
H_{c}(t) & = & \sum_{k=1}^{3}\left[h^{k}(t)\sigma_{k}\otimes I+h^{k+3}(t)I\otimes\sigma_{k}\right]\label{Hc1}
\end{eqnarray}
Our goal in this second stage of the analysis is to find the optimal
$h^{k}(t),$ for
$k=1,\ldots,6$ such that the functional~\cite{PhysRevA.110.042601} \begin{widetext}
\begin{eqnarray}\label{Jfunctional}
\mathcal{J}(H,U,\Lambda) & = & \int_{0}^{\tau}\mathrm{d}t\,\frac{1}{4}\mathrm{Tr}\mleft[\frac{1}{2}H_{c}(t)H_{c}(t)+\Lambda(t)\left(i\frac{\mathrm{d}U(t)}{\mathrm{d}t}U^{\dagger}(t)-H_\mathrm{opt}(t)\right)\mright]\label{Func}
\end{eqnarray}
\end{widetext}is minimal, where $\tau$ is the time interval corresponding to
the duration of the gate operation, $\Lambda(t)$ is the
co-state (the Hermitian matrix containing the instantaneous Lagrange
multipliers), and we use $H(t),$ $U(t),$
and $\Lambda(t)$ as the functions to be varied independently
to minimize the functional $\mathcal{J}(H,U,\Lambda),$ which is
equivalent to satisfying Pontryagin's Maximum Principle~\cite{PRXQuantum.2.030203}. Minimization of this functional implies energy optimization. The reason is that the quantity
\begin{eqnarray}\label{Efunctional}
    E(H_c) & = & \int_0^\tau \mathrm{d}t \, \frac{1}{4} \mathrm{Tr}\mleft[ \frac{1}{2} H_c(t) H_c(t) \mright]
\end{eqnarray}
is minimal when the eigenstates of $H_c(t)$ are minimal for all $t \in [0,\tau]$. The remaining term in the integrand of Eq.~(\ref{Func}) is the constraint imposed by the Schr\"odinger equation, which is the relation that $U(t)$ and $H_\mathrm{opt}(t)$ must obey.

As explained in~\cite{Swaddle-dissertation,SWADDLE20173391,Perrier_2020,PhysRevA.110.042601,dasilva2025montecarloapproachfinding}, from Eq.~(\ref{Func}) we can derive the geodesic equations:
\begin{eqnarray}
i\frac{\mathrm{d}U(t)}{\mathrm{d}t} & = & H_\mathrm{opt}(t)U(t), \label{Schr}
\end{eqnarray}
\begin{eqnarray}
i\frac{\mathrm{d}\Lambda(t)}{\mathrm{d}t} & = & \left[H_\mathrm{opt}(t),\Lambda(t)\right],  \label{adjL}
\end{eqnarray}
and
\begin{eqnarray}
\mathcal{P}\mleft[\Lambda(t)\mright] & = & H_{c}(t), \label{Hc}
\end{eqnarray}
where $\mathcal{P}$ is the projector onto the sub-Riemannian~\cite{montgomery2002tour}
subspace spanned by the distribution $\Delta,$ defined by
\begin{eqnarray}\label{distribution}
\Delta & = & \left\{ \sigma_{k}\otimes I,I\otimes\sigma_{k}|k=1,2,3\right\} .
\end{eqnarray}
As detailed in~\cite{PhysRevA.110.042601}, we must find $\Lambda(0)$
such that $U(\tau)$ is the closest we can get to the target
$U_\mathrm{target}$.

\subsection{Adjoint Sensitivity Analysis}\label{subsec:adjoint_analysis}

The geodesic equations, Eqs.~(\ref{Schr}), (\ref{adjL}), and (\ref{Hc}), are obtained by minimizing the control Hamiltonian, $H_{c}(t).$ To derive these equations, we do not minimize the infidelity to obtain, at $t=\tau,$ the target unitary operator.  We can, however, reformulate these three geodesic equations into a single non-linear Schrödinger equation~\cite{SWADDLE20173391}:
\begin{eqnarray}
    i\frac{\mathrm{d}}{\mathrm{d}t}U(t) & = & \left\{ H_{d} + \mathcal{P}\mleft[U(t)\Lambda(0)U^{\dagger}(t)\mright]\right\}U(t), \label{nlSchr}
\end{eqnarray}
as explained in detail in~\cite{Swaddle-dissertation,SWADDLE20173391,Perrier_2020,PhysRevA.110.042601,dasilva2025montecarloapproachfinding}. For each choice of initial adjoint operator, $\Lambda(0),$ by integrating Eq.~(\ref{nlSchr}) forward in time, we obtain a geodesic that ends up with a corresponding operator $U(\tau),$ whose fidelity with $U_\mathrm{target}$ has not been optimized. As we shall see in the next section, we can use numerical methods to find the optimal $\Lambda(0)$ that maximizes the fidelity between $U(\tau)$ and $U_\mathrm{target}.$ Here, however, we introduce a second variational principle to minimize the infidelity resulting from a given choice of $\Lambda(0).$

We define the infidelity we want to minimize as given by
\begin{eqnarray}\label{fidelity_metric}
\mathcal{I}(U) & = & 1 - \left|\frac{1}{4}\mathrm{Tr}\mleft[U_\mathrm{target}^{\dagger}U(\tau)\mright]\right|^{2},\label{infidel}
\end{eqnarray}
which is $1$ minus the fidelity, Eq. (\ref{fidelity-CDD}).
We are looking for a unitary operator $U(t),$ whose terminal value, at $t=\tau,$ minimizes $\mathcal{I}(U),$ Eq.~(\ref{infidel}). We must look for the desired $U(t)$ only among those that solve Eq.~(\ref{nlSchr}), given the possible adjoint matrices $\Lambda(0).$ Thus, the minimization of Eq.~(\ref{infidel}) is constrained by Eq.~(\ref{nlSchr}), which warrants the introduction of new Lagrange multipliers. Accordingly, we define the functional:
\begin{widetext}
    \begin{eqnarray}
    \mathcal{K}\mleft(U,\Gamma,\Lambda(0)\mright) & = & \mathcal{I}(U)+\int_{0}^{\tau}\mathrm{d}t\,\frac{1}{4}\mathrm{Tr}\mleft\{\Gamma(t)\left[i\frac{\mathrm{d}U(t)}{\mathrm{d}t} U^{\dagger}(t) - H_{d} - \mathcal{P}\left(U(t)\Lambda(0)U^{\dagger}(t)\right) \right]\mright\}. \label{Kfunctional}
\end{eqnarray}
\end{widetext}
We emphasize that the functional $\mathcal{K}\mleft(U,\Gamma,\Lambda(0)\mright)$ depends on $U(t),$ $\Gamma(t),$ and $\Lambda(0),$ taken as mutually independent variables, which is why we introduce the Lagrange multipliers. Although we have $U(t)$ and $\Gamma(t)$ as independent variables at each time $t\in\left[0,\tau\right],$ $\Lambda(0)$ is an independent variable given only for a single instant, $t=0.$
We now proceed with the usual variational calculus. By imposing that $\delta \mathcal{K}$ is zero for all variations of $\Gamma(t),$ we immediately re-obtain Eq.~(\ref{nlSchr}). The careful analysis of the variations of $U(t),$ also taking into account the terminal point at $t=\tau,$ gives a terminal condition for $\Gamma(\tau),$
\begin{eqnarray}
\Gamma(\tau) & = & \frac{i}{16}\mathrm{Tr}\mleft[U_\mathrm{target}^{\dagger}U(\tau)\mright]U_\mathrm{target}U^{\dagger}(\tau) \nonumber \\
& - & \frac{i}{16}\mathrm{Tr}\mleft[U^{\dagger}(\tau)U_\mathrm{target}\mright]U(\tau)U_\mathrm{target}^{\dagger}, \label{termcond}
\end{eqnarray}
and an adjoint equation for $\Gamma(t),$
\begin{eqnarray}
i\frac{\mathrm{d}\Gamma(t)}{\mathrm{d}t} & = & \left[H_{d}+\mathcal{P}\mleft[U(t)\Lambda(0)U^{\dagger}(t)\mright],\Gamma(t)\right] \nonumber \\ & + & \left[\mathcal{P}\mleft[\Gamma(t)\mright],U(t)\Lambda(0)U^{\dagger}(t)\right]. \label{adj}
\end{eqnarray}
Finally, we vary $\Lambda(0)$ and, by imposing $\delta \mathcal{K} =0 $ for all variations $\delta\Lambda(0)$, we get the following condition:
\begin{eqnarray}
\int_{0}^{\tau}\mathrm{d}t\,U^{\dagger}(t)\mathcal{P}\mleft[\Gamma(t)\mright]U(t) & = & 0.
\end{eqnarray}
This is not a useful equation, as it implies that we must continually change $\Lambda(0)$ until this condition is satisfied.

Instead, we can think of the functional derivative of $\mathcal{K}$, Eq.~(\ref{Kfunctional}), with respect to $\Lambda(0)$
before imposing that it is zero. Therefore, we use a matrix basis $\left\{\alpha_{k}\right\}$ and write
\begin{eqnarray}
\Lambda(0) & = & \sum_{k=1}^{D}\alpha_{k}\lambda_{k},\label{lambdak}
\end{eqnarray}
where $\lambda_{k}$ are real parameters given by
\begin{eqnarray}
\lambda_{k} & = & \frac{1}{4}\mathrm{Tr}\mleft[\alpha_{k}\Lambda(0)\mright].
\end{eqnarray}
Hence, taking variations, we obtain
\begin{eqnarray}
\delta\lambda_{k} & = & \frac{1}{4}\mathrm{Tr}\mleft[\alpha_{k}\delta\Lambda(0)\mright].
\end{eqnarray}
It is now operationally more manageable to define the gradient components,
\begin{eqnarray}
\frac{\delta \mathcal{K}}{\delta\lambda_{k}} & = & -\int_{0}^{\tau}\mathrm{d}t\,\mathrm{Tr}\mleft\{ \alpha_{k}U^{\dagger}(t)\mathcal{P}\mleft[\Gamma(t)\mright]U(t)\mright\} . \label{grad}
\end{eqnarray}

\subsubsection{Learning $\Lambda(0)$}
We use Eq.~(\ref{grad}) as part of the algorithm with which we learn the optimal $\Lambda(0).$ Given a real parameter $\eta,$ where $0<\eta<1,$ the learning rate, we iteratively correct $\Lambda(0)$ using the procedure:
\begin{eqnarray}
\Lambda(0) & \rightarrow & \Lambda(0)-\eta\sum_{k=1}^{D}\alpha_{k}\frac{\delta \mathcal{K}}{\delta\lambda_{k}}, \label{Lambda_variation}
\end{eqnarray}
where we use Eqs.~(\ref{lambdak}) and (\ref{grad}).

\subsubsection{The algorithm for learning $\Lambda(0)$}
\begin{enumerate}
    \item Given a choice of $\boldsymbol{\lambda}\in\mathbb{R}^{D},$
we calculate $\Lambda(0)$ using Eq.~(\ref{lambdak}).
    \item Then, we calculate $U(t)$
using Eq.~(\ref{nlSchr}), from $t=0$ to $t=\tau.$
    \item  Next, we take $U(\tau)$ and
calculate $\Gamma(\tau)$ using the terminal condition, Eq.~(\ref{termcond}).
    \item After this, we propagate $\Gamma(t)$ backward in time,
from $t=\tau$ to $t=0,$ by integrating Eq.~(\ref{adj}) since we already have $U(t),$ obtained from the previous
integration of the non-linear Schrödinger equation, Eq.~(\ref{nlSchr}), forward in time.
    \item Then, having also obtained $\Gamma(t),$ we can calculate
the integral
\begin{eqnarray}
\mathcal{F}\mleft(\boldsymbol{\lambda}\mright) & = & -\int_{0}^{\tau}\mathrm{d}t\,\mathrm{Tr}\mleft\{ U^{\dagger}(t)\mathcal{P}\mleft[\Gamma(t)\mright]U(t)\Lambda(0)\mright\} , \label{calF}
\end{eqnarray}
whose gradient is given by Eq.~(\ref{grad}).
    \item At this point, we compute the infidelity, Eq.~(\ref{infidel}), with the new $\Lambda(0)$ and decide whether to re-iterate this algorithm starting from Step 3.
\end{enumerate}

\subsubsection{How this algorithm works}
We can see why the algorithm given above works by the following argument. The corresponding change in $\mathcal{F}\mleft(\boldsymbol{\lambda}\mright)$
becomes, using Eqs.~(\ref{lambdak}), (\ref{grad}) and (\ref{calF}),
\begin{eqnarray}
\delta\mathcal{F}\mleft(\boldsymbol{\lambda}\mright) & = & -\eta\sum_{k=1}^{D}\left[\frac{\partial\mathcal{F}\mleft(\boldsymbol{\lambda}\mright)}{\partial\lambda_{k}}\right]^{2}, \label{deltaF}
\end{eqnarray}
which is less than zero. Given a solution of Eq.~(\ref{nlSchr}), the functional of Eq.~(\ref{Kfunctional}), $\mathcal{K}$, equals the infidelity, Eq.~(\ref{infidel}), calculated for a given $\Lambda(0)$ and its corresponding $U(\tau),$ obtained via forward integration of Eq.~(\ref{nlSchr}). If then we introduce a variation of such a $\mathcal{K}$ as given above, Eq.~(\ref{Lambda_variation}), we obtain
\begin{eqnarray}
\delta \mathcal{K} & = & \delta\mathcal{F}\mleft(\boldsymbol{\lambda}\mright) \nonumber \\
 & = & -\eta\sum_{k=1}^{D}\left[\frac{\partial\mathcal{F}\mleft(\boldsymbol{\lambda}\mright)}{\partial\lambda_{k}}\right]^{2}<0,
\end{eqnarray}
where we used Eq.~(\ref{deltaF}), and $\mathcal{K}$ decreases. This result means that the infidelity decreases as desired.

\subsection{Other Numerical Solution Strategies}

The search for the costate $\Lambda(0)$ that minimizes infidelity is generally hard~\cite{PhysRevLett.114.170501,PhysRevA.110.042601}. Here we present two additional techniques~\cite{dasilva2025montecarloapproachfinding}, besides the one presented in the last section, which result in optimal $\Lambda(0)$:
a newly-developed Monte Carlo approach and an optimal Krotov method. We compare these two methodologies
and discuss the advantages and disadvantages of each in the context
of our analysis. But similarly to the presented method, the main objective is to minimize the infidelity function given by Eq.~(\ref{infidel}).
% \begin{eqnarray}\label{fidelity_metric}
% \mathcal{F}_{U} & = & \frac{1}{4}\left|\mathrm{Tr}\left[U_\mathrm{CZ}^{\dagger}U(\tau)\right]\right|.\label{Ufid}
% \end{eqnarray}

\subsubsection{Monte Carlo}

This numerical approach is based on the idea of treating Eq.~(\ref{nlSchr}) as a map $\mathbb{R}^{D} \rightarrow \mathbb{R}^{D}$.
From Eq.~(\ref{lambdak}) it is clear how $\Lambda(0)$ can be identified with an element of $\mathbb{R}^D$. Upon choosing some $\Lambda(0)$, Eq.~(\ref{nlSchr}) can be numerically solved for the time interval $[0, \tau]$, and the resulting $U(\tau)$ can be written in the form
\begin{eqnarray}\label{utau}
    U(\tau) & = & \exp\mleft[-i \sum_{k=1}^D c_k \alpha_k \mright].
\end{eqnarray}
Hence, $U(\tau)$ can also be identified by the $D$ real numbers $c_k$. Essentially, solving the geodesic equation can be thought of as a map that takes $D$ numbers $\lambda_k$, and outputs a unitary matrix that can be described by another set of $D$ numbers $c_k$, that is, $\mathbb{R}^{D} \rightarrow \mathbb{R}^{D}$. Since the $c_k$ coefficients appear inside a complex exponential function, which is not uniquely invertible, the costate $\Lambda(0)$ that leads to some $U(\tau)$ is not unique. Furthermore, since Eq.~(\ref{nlSchr}) depends on the projection $\mathcal{P}$, which is not invertible, it is not possible to determine a $\Lambda(0)$ by solving the equation directly backwards in time.

With these points considered, the Monte Carlo method consists of a random sampling of $D$-dimensional real arrays that serve as $\Lambda(0)$ in Eq.~(\ref{nlSchr}). Then the equation is numerically solved for each array, and the resulting $U(\tau)$ is stored as a $D$-dimensional array of numbers $c_k$ obtained through inversion of Eq.~(\ref{utau}), always considering the main logarithmic branch. Then, given a desired two-qubit gate $U_\mathrm{target}$, it is also associated with a $D$-dimensional set of numbers, analogous to the relation of Eq.~(\ref{utau}). Consider that such a set of numbers is given by $\left\{\widetilde{c}_k\right\}$. Using the criterion 
\begin{eqnarray}
    \underset{\{c_k\}}{\mathrm{min}} \mleft[ \sum_{k=1}^D \left| c_k - \widetilde{c}_k \right| \mright],
\end{eqnarray}
We determine which set of numbers generates the unitary matrix closest to the desired one, and the specific $\Lambda(0)$ that leads to that unitary matrix is used as a candidate for generating the optimal control producing $U_\mathrm{target}$. Let $\widetilde{\Lambda}(0)$ represent such a co-state. We feed $\widetilde{\Lambda}(0)$, Eq.~(\ref{nlSchr}), and the gate infidelity function of Eq.~(\ref{infidel}) to a minimization function, such as Mathematica Wolfram's \texttt{FindMinimum}, to vary $\widetilde{\Lambda}(0)$ in the geodesic equation until the resultant $U(\tau)$ is the closest possible to $U_\mathrm{target}$, meaning $\mathcal{I}(U_\mathrm{target}) \rightarrow 0$.

This process works not only for two-qubit gates, but also for single-qubit gates such as $U \otimes I$ and $I \otimes U$, where $I$ is the single-qubit space identity operator. This method was introduced in~\cite{dasilva2025montecarloapproachfinding}, where its advantages and disadvantages are explained, as are all its details.

\subsubsection{Krotov}

We also employ the well-known Krotov method~\cite{PhysRevA.68.062308,10.21468/SciPostPhys.7.6.080,Fernandes_2023}, where the optimization of $H_c(t)$ is performed to achieve the desired quantum gate $U_\mathrm{target}$ without the constraint of minimizing the energy. Numerically, the Krotov method depends on an initial guess for $h^{k}(t)$  in Eq.~(\ref{Hc1}). In this case, the strategy for achieving controls with small energy is to start the self-consistent calculations with constant initial guesses having small amplitudes.

% \subsection{Fidelity Metric}

% The criterion to find the correct $\Lambda(0),$ leading
% to the desired target $U_\mathrm{CZ},$ is to maximize the unitary-operator
% fidelity that we define by
% \begin{eqnarray}\label{fidelity_metric}
% \mathcal{F}_{U} & = & \frac{1}{4}\left|\mathrm{Tr}\left[U_\mathrm{CZ}^{\dagger}U(\tau)\right]\right|.\label{Ufid}
% \end{eqnarray}

\section{Results}

% \textbf{\textcolor{red}{(Esta seção foi sugerida como template pelo
% ChatGPT4o)}}In this section, we summarize the numerical results obtained
% from the two-stage control method combining Continuous Dynamical Decoupling
% (CDD) and geodesic-informed optimal control.

% The key performance indicators are:

% - Fidelity of the CZ gate operation under the combined method.

% - Robustness against environmental noise modeled as a Redfield decoherence
% process.

% - Robustness against variations in the native Hamiltonian couplings
% (i.e., fabrication imperfections and crosstalk).

% \#\#\# 5.1 CDD Performance

% - Fidelity benchmarks as a function of CDD frequency $\omega$
% (summarize results from Section 3.2).

% - Validation that the effective drift Hamiltonian Hd accurately describes
% the dynamics (summarize results from Section 3.3).

% \#\#\# 5.2 Optimal Control Performance

% - Final fidelity achieved by the CZ gate using each numerical optimization
% method (Krotov, Monte Carlo, PINN).

% - Compare convergence performance and computational effort.

% \#\#\# 5.3 Robustness Analysis

% - Sensitivity of the optimized gate to residual noise or drift in
% the Hamiltonian.

% - (Optional) Any analysis of how the gate behaves under different
% levels of decoherence or parameter fluctuations.

\subsection{CZ gate}

As a first test for the three methods, we chose the two-qubit CZ gate, which we define as
\begin{align}\label{UCZ}
    U_\mathrm{CZ} &\equiv \exp\mleft(-i\frac{\pi}{4}\mright)\begin{pmatrix}
        1 & 0 & 0 & 0 \\ 0 & 1 & 0 & 0 \\ 0 & 0 & 1 & 0 \\ 0 & 0 & 0 & -1
    \end{pmatrix}
\end{align}
The phase factor $\exp\left(-i\pi/4\right)$ is just to ensure a unitary determinant for the resulting matrix, resulting in an element of $\mathrm{SU}(4)$. Nevertheless, a global phase factor is irrelevant to gate fidelity, Eq.~(\ref{infidel}). With this matrix representation, we are automatically selecting a two-qubit basis ordered as $\ket{0,0}, \ket{0,1}, \ket{1,0}, \ket{1,1}$. It is direct to verify that the Hamiltonian
\begin{align}
    H_\mathrm{CZ} \equiv \frac{\pi}{4\tau} \left( I \otimes \sigma_z + \sigma_z \otimes I - \sigma_z \otimes \sigma_z \right)
\end{align}
results in $U_\mathrm{CZ}$ at $t=\tau$. This is different from Eq.~(\ref{Hd}), which contains a term proportional to $\sigma_y \otimes \sigma_y$, and $J_{3,3}$ is different from $\pi/(4\tau)$ in general. The objective then is to determine the single-qubit control fields that will modify the total Hamiltonian such that $U_\mathrm{CZ}$ is reached at $t=\tau$ with high fidelity, Eq.~(\ref{infidel}).
For this case, $U_\mathrm{target}$ corresponds to Eq.~(\ref{UCZ}).

\begin{figure}
    \centering
    \includegraphics[width=0.47\textwidth]{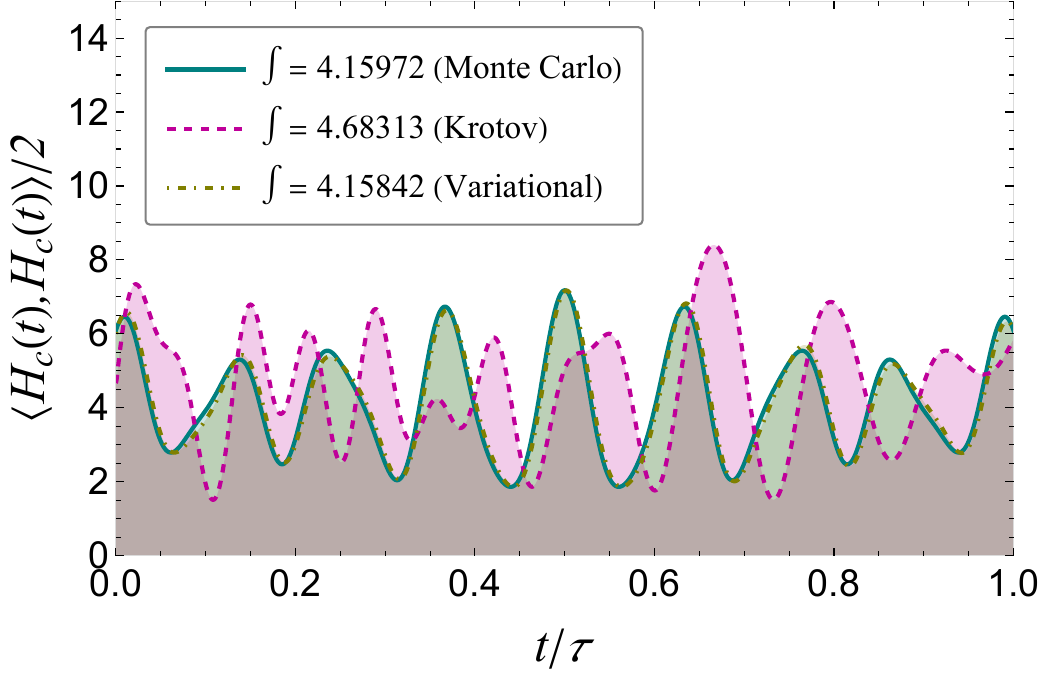}
    \includegraphics[width=0.47\textwidth]{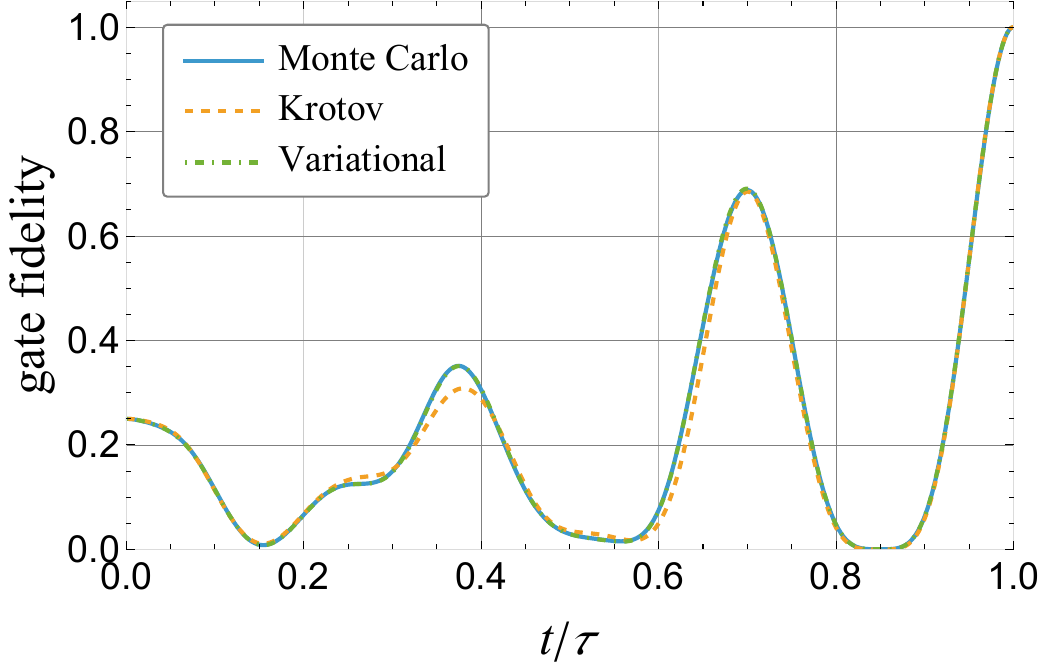}
    \caption{At the top is shown the integrand of the energetic cost functional, Eq.~(\ref{Efunctional}). The area under each curve corresponds to the actual cost of implementing the gate. In the legends, we show the numerical value of the integral for each curve, which is in units of $\hbar^2/\tau$. At the bottom is the gate fidelity as a function of time. Again, the curves for the Monte Carlo and Variational methods are practically the same.}
    \label{CZ_graphs}
\end{figure}

\begin{figure}
    \centering
    \subfloat[\label{CZcontrol-MonteCarlo}Monte Carlo]{%
    \includegraphics[width=1\columnwidth]{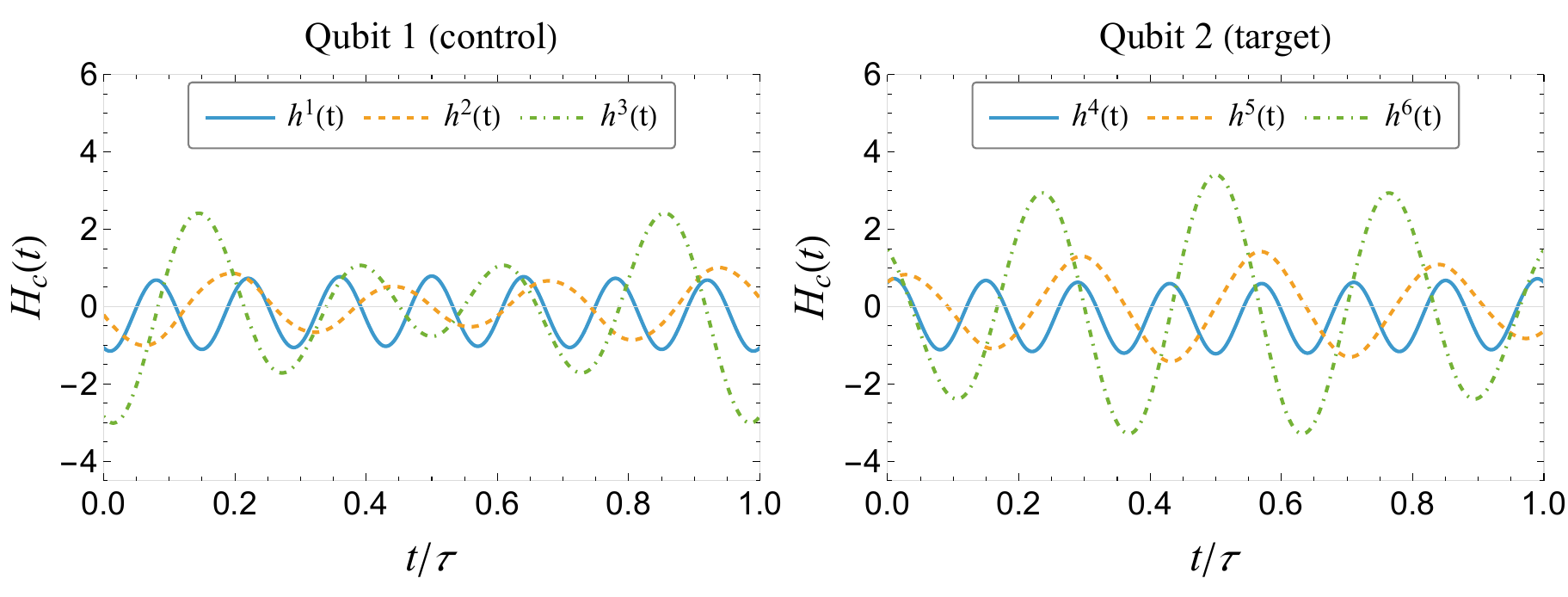}%
    } \\ \vspace{-0.3cm}
    \subfloat[\label{CZcontrol-Krotov}Krotov]{%
    \includegraphics[width=1\columnwidth]{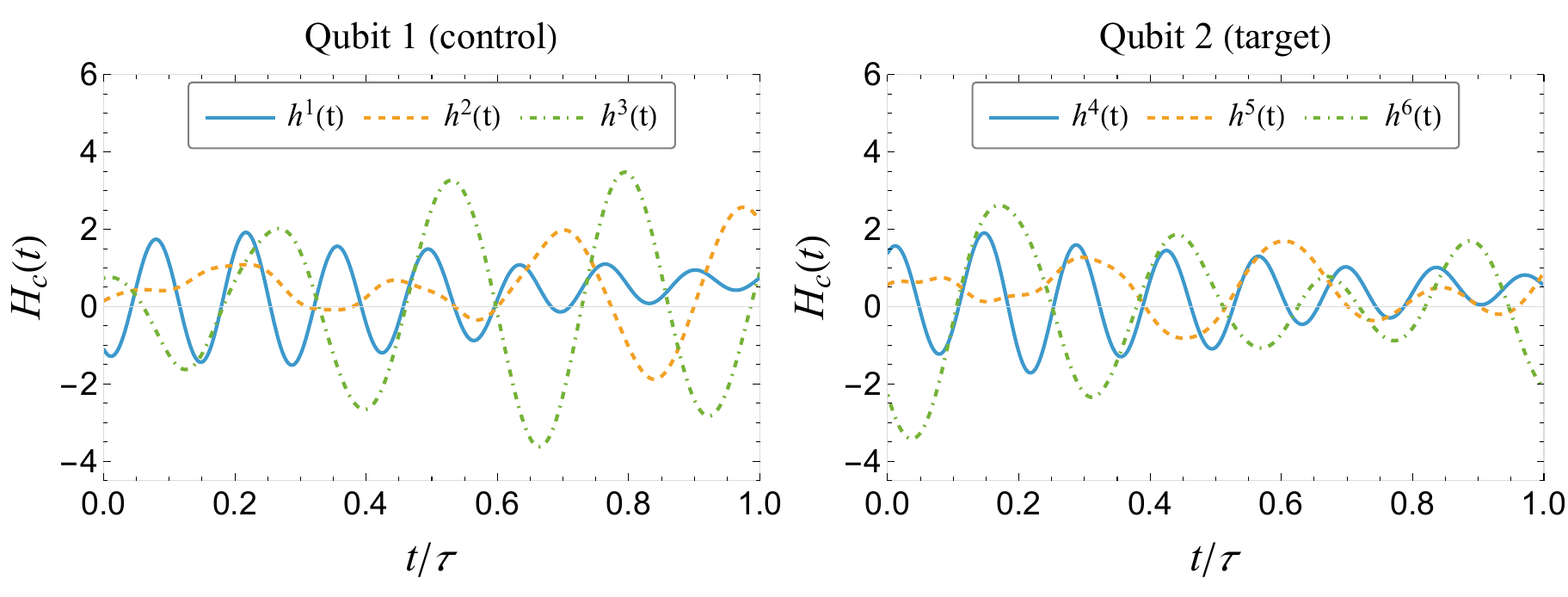}%
    } \\ \vspace{-0.3cm}
    \subfloat[\label{CZcontrol-Variational}Variational]{%
    \includegraphics[width=1\columnwidth]{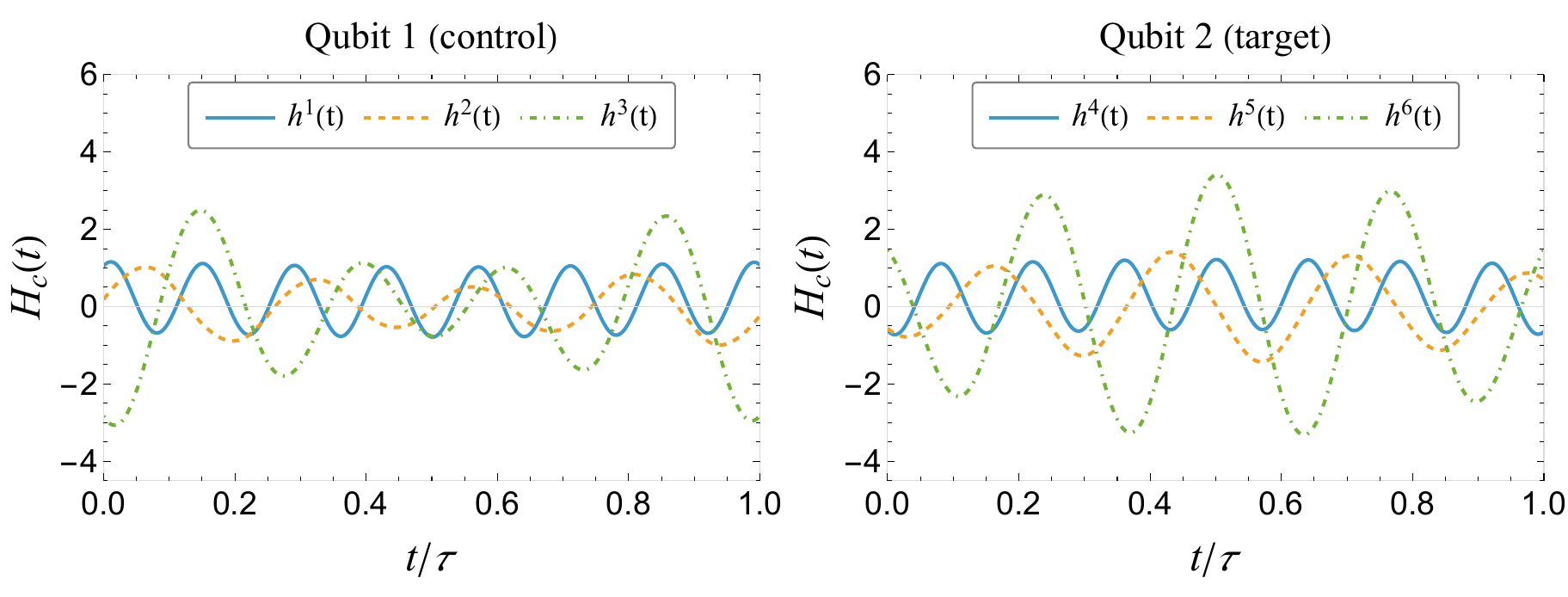}%
    }
    \caption{Optimal control Hamiltonian for the CZ gate obtained with each of the methods: (a) Monte Carlo, (b) Krotov, and (c) Variational. The indices $j=1,2,3$ correspond to the operators $\sigma_j \otimes I$, which operate on the first qubit, and $j=4,5,6$ correspond to the operators $I \otimes \sigma_{j-3},$ as explained just following Eq. (\ref{Hc1}). Notice that the solutions in (a) and (c) differ only by a multiplicative factor of $-1$ in the $\sigma_x$ and $\sigma_y$ components.}
    \label{control_fields}
\end{figure}

This was done for the three methods mentioned in this work: the Monte Carlo, Krotov, and Variational methods, thoroughly explained in Sec.~\ref{subsec:adjoint_analysis}. The results for the gate fidelity as a function of time, as well as the energetic cost functional, are shown in Fig.~\ref{CZ_graphs}.
Notice that for the Variational and Monte Carlo methods, the solutions appear to be identical. This is expected since both are rooted in the principle of minimizing the functional of Eq.~(\ref{Jfunctional}), their difference lies only in the way of determining the initial co-state for the geodesic equation. However, in Fig.~\ref{control_fields} the three components of the control Hamiltonian are plotted for each qubit of the pair, and, as can be seen, although the obtained components $\sigma_z$ ($h^3(t)$ and $h^6(t)$) are the same for both methods, the components $\sigma_x$ and $\sigma_y$ are multiplied by $-1$ between them. From Fig.~\ref{CZ_graphs} we conclude that this difference has no impact on gate fidelity or the energetic cost of implementation.

Because the CZ gate is symmetrical, it is also important to note that exchanging the control fields between the qubits also results in the CZ gate with the same fidelity. The choice to label the first qubit as the control and the second as the target was just an adopted convention for this work.

\subsection{Other gates}

We also tested the methods for two other two-qubit gates. Of these additional gates, the first chosen was the CX (controlled-not) gate, which is defined as
\begin{align}\label{UCX}
    U_\mathrm{CX} &\equiv \exp\mleft(-i\frac{\pi}{4}\mright)\begin{pmatrix}
        1 & 0 & 0 & 0 \\ 0 & 1 & 0 & 0 \\ 0 & 0 & 0 & 1 \\ 0 & 0 & 1 & 0
    \end{pmatrix}.
\end{align}
This operation generates a Bell state when applied to the product state $\ket{+,0}$, where $\ket{+} = (\ket{0}+\ket{1})/\sqrt{2}$. To further test the capabilities of the optimal control methods, we selected the second additional two-qubit gate to be
\begin{align}\label{generic_U}
    U_\mathrm{R} \equiv (U_1 \otimes U_2) U_\mathrm{CX},
\end{align}
that is, the CX gate followed by local qubit rotations, where $U_1$ and $U_2$ are two Haar-random single-qubit unitaries given by
\begin{align*}
    U_1 &\approx \begin{pmatrix}
        -0.942908-0.158967i & 0.0207764 +0.291929i \\ 
        -0.0207764+0.291929i & -0.942908+0.158967i
    \end{pmatrix}
\end{align*}
and
\begin{align*}
    U_2 &\approx \begin{pmatrix}
        0.260618 +0.772926i & 0.532428 +0.226236i \\ 
        -0.532428+0.226236i & 0.260618 -0.772926i
    \end{pmatrix},
\end{align*}
respectively~\cite{Mezzadri2007}.

Equation~(\ref{generic_U}) can be thought of as a generic entangling gate that is locally equivalent to a CX (and CZ) gate. The motivation is that, if an optimal control for such a gate is possible to obtain as a protocol for external fields acting within the fixed gate time $\tau$, this can significantly decrease depth in circuits since any entangling gate locally equivalent to CX followed and/or preceded by local operations could always be treated as a single two-qubit gate.

\begin{figure}[h]
    \centering
    \includegraphics[width=0.47\textwidth]{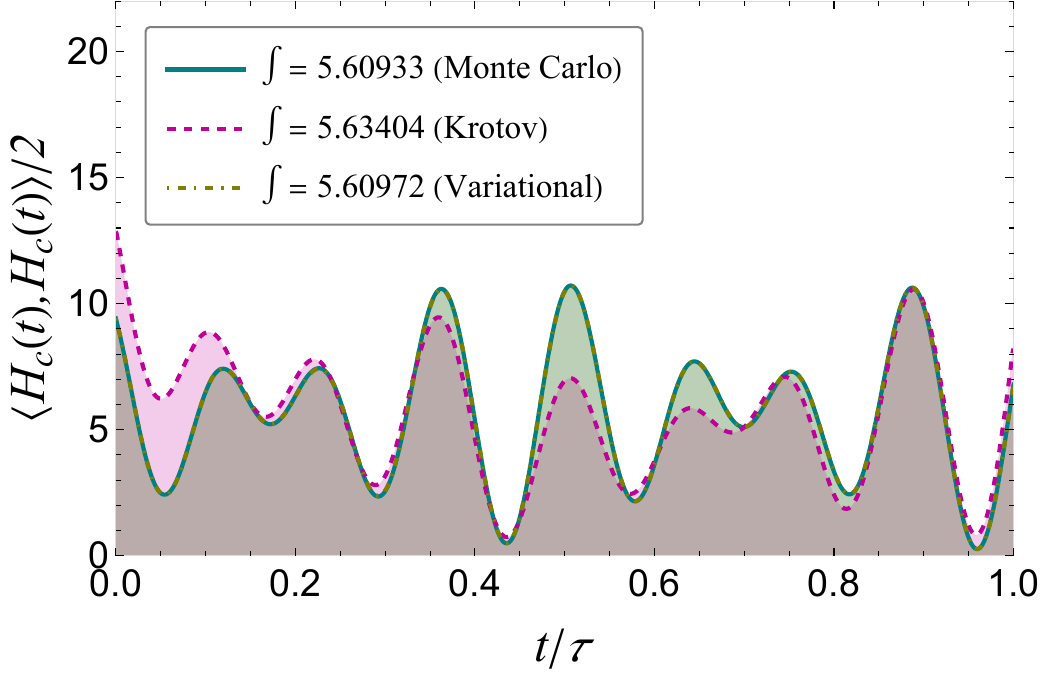}
    \includegraphics[width=0.47\textwidth]{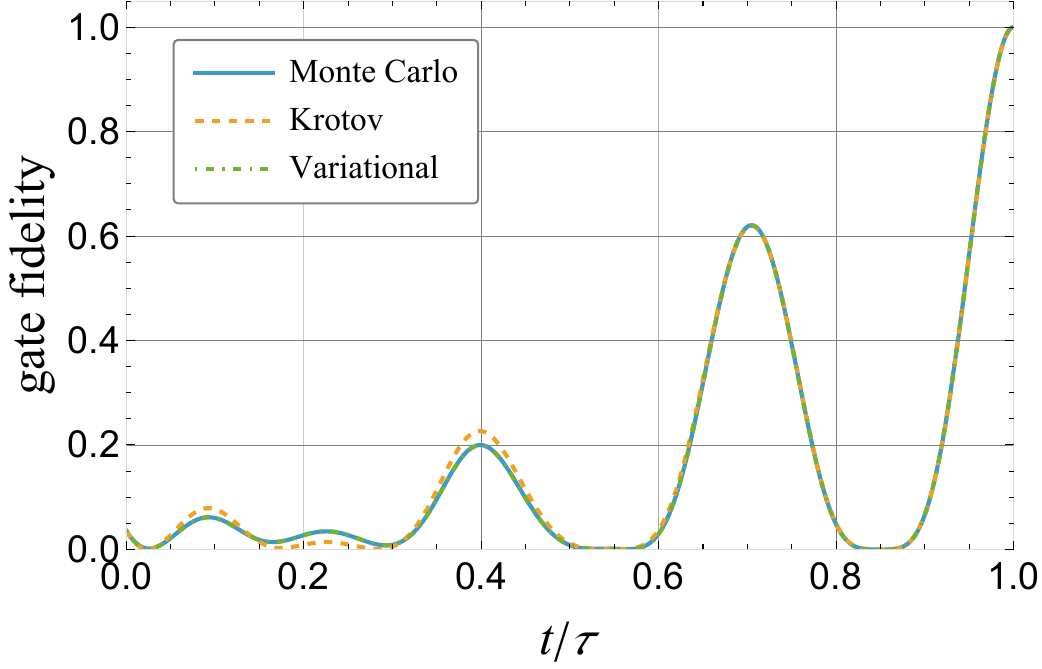}
    \caption{Analogous to Fig.~\ref{CZ_graphs}, but for the general entangling gate $U_\mathrm{R}$ given by Eq.~(\ref{generic_U}), which is locally equivalent to the CX (and CZ) gate, meaning they only differ by local operations. At the top is shown the integrand of the energetic cost functional, Eq.~(\ref{Efunctional}). In the legends, we show the numerical value of the integral for each curve, which is in units of $\hbar^2/\tau$. At the bottom is the gate fidelity as a function of time.}
    \label{U1U2CX_graphs}
\end{figure}

Similarly to Fig.~\ref{CZ_graphs}, the results for the general entangling gate $U_\mathrm{R}$ given by Eq.~(\ref{generic_U}) are shown in Fig.~\ref{U1U2CX_graphs}. Again, the two methods based on sub-Riemannian geodesics reach the same result for gate fidelity and energetic cost.

\begin{comment}

\begin{figure}[h!]
    \centering
    \includegraphics[width=0.47\textwidth]{figures/CX_en.pdf}
    \includegraphics[width=0.47\textwidth]{figures/CX_fid.pdf}
    \caption{Analogously to Fig.~\ref{CZ_graphs}, but for the $U_\mathrm{CX}$ entangling gate, given by Eq.~(\ref{UCX}), at the top, it is shown integrand of the energetic cost functional, Eq.~(\ref{Efunctional}). In the legends, we show the numerical value of the integral for each curve, which is in units of $\hbar^2/\tau$. At the bottom is the gate fidelity as a function of time.}
    \label{CX_graphs}
\end{figure}
\end{comment}

\subsection{Constraints on the control subalgebra}

One possible caveat of the presented method is that it assumes complete and arbitrary control over single-qubit rotations for any of the three axes in the Bloch sphere. This is generally not true for physical platforms such as those involving superconductors, whose qubits are arranged in a two-dimensional lattice, and trapped atoms, where $\sigma_z$ pulses are often applied virtually through pulses along the $\sigma_x$ and $\sigma_y$ directions.
One way to address this issue is to remove one of the components in the spanning set of the distribution given by Eq.~(\ref{distribution}). Suppressing one of the indices will imply that one of the directions will only be achieved through commutation relations between the other two. With this in mind, the optimal control was re-obtained for $U_\mathrm{CX}$ and $U_\mathrm{R}$, of Eqs.~(\ref{UCX}) and (\ref{generic_U}), despite individually suppressing each component from the control operations for each case.

\begin{table}[h]
    \centering
\begin{tabular}{|c|c|c|c|c|}
\hline 
\multicolumn{2}{|c|}{} & \multicolumn{3}{c|}{Energy cost ($\hbar^{2}/\tau$)}\tabularnewline
\hline 
\hline 
Gate & Control directions & Monte Carlo & Krotov & Variational\tabularnewline
\hline 
CX & $x,y,z$ & 3.41619 & 3.49247 & 3.41619\tabularnewline
\hline 
CX & $y,z$ & 6.48627 & 8.26504 & 6.48769\tabularnewline
\hline 
CX & $x,z$ & 5.6051 & 5.7918 & 5.60339\tabularnewline
\hline 
CX & $x,y$ & 5.33592 & 5.4952 & 5.33736\tabularnewline
\hline 
R & $x,y,z$ & 5.60933 & 5.63404 & 5.60972\tabularnewline
\hline 
R & $y,z$ & 5.74649 & 6.01273 & 5.74939\tabularnewline
\hline 
R & $x,z$ & 8.43324 & 11.1651 & 8.43734\tabularnewline
\hline 
R & $x,y$ & 8.20955 & 11.9895 & 8.20941\tabularnewline
\hline 
\end{tabular}
\caption{Results for energy cost, in units of $\hbar ^{2}/\tau ,$ for the CX and R gates of Eqs.~(\ref{UCX}) and (\ref{generic_U}), respectively. We show the costs using the different methods, Monte Carlo, Krotov, and variational, for each case of control restriction. Thus, we present the unrestricted case, where the control is allowed in all three spatial directions, as well as when one of the directions is unavailable for control in turn.}
\label{tab:tab1}
\end{table}
The results are shown in Table~\ref{tab:tab1}. For all three cases and for each quantum gate, it is possible to achieve unitary values of gate fidelity. This suggests that the methods described in this work can be useful in realistic experimental scenarios involving control of only two axes in the Bloch sphere.

\begin{comment}

\begin{figure}[h]
    \centering
    \includegraphics[width=0.97\columnwidth]{figures/CX_en_ModControl.pdf}
    \includegraphics[width=0.97\columnwidth]{figures/CX_fid_ModControl.pdf}
    \caption{Energetic cost and gate fidelity for $U_\mathrm{CX}$ gate with each of the three components separately suppressed. In all cases, the three methods managed to find an optimal control Hamiltonian that results in $U_\mathrm{CX}$ with unitary fidelity at $t=\tau$. Similar to the previous results, the Variational and Monte Carlo methods reached equivalent solutions.}
    \label{CXsuppressed_graphs}
\end{figure}

\end{comment}

\begin{widetext}

\begin{figure}[h!]
    \centering
    \includegraphics[width=0.97\columnwidth]{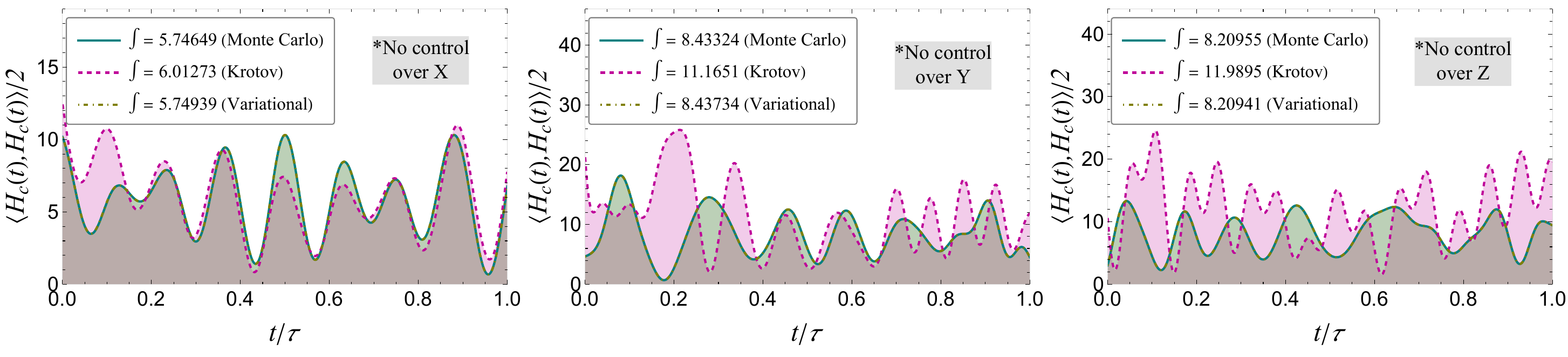}
    \includegraphics[width=0.97\columnwidth]{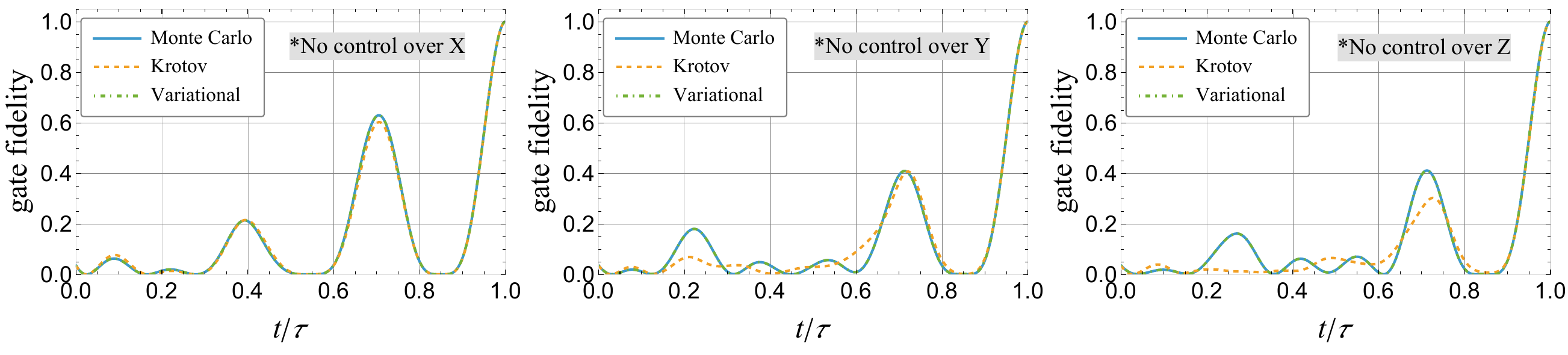}
    \caption{Energetic cost and gate fidelity for $U_\mathrm{R}$ gate with each of the three components separately suppressed. In all cases, the three methods can find an optimal control Hamiltonian that results in $U_\mathrm{R}$ with unitary fidelity at $t=\tau$. Similar to the previous results, the Variational and Monte Carlo methods reached equivalent solutions.}
    \label{U1U2Xsuppressed_graphs}
\end{figure}

\end{widetext}

\section{Discussion of the results}

The results obtained through the three optimization strategies, the Monte Carlo costate sampling, the geodesic variational approach, and the Krotov method pinpoint several important features of our proposed two-stage control scheme, which combines continuous dynamical decoupling (CDD) with variational optimal control.

The first observation is that mitigation of native Hamiltonian imperfections using CDD is effective and essential for the next phase of the optimization protocol. The application of the CDD technique eliminates all residual couplings $J_{\mu,\nu}$ but the intended $J_{3,3}$ term. We demonstrated numerically that even moderate 2-20 GHz microwave-frequency fields suppress unwanted dynamics, with fidelities better than 0.99999. Thus, we have shown that the drift Hamiltonian can be perfected to a high degree before optimizing the desired gate. These results are relevant in the context of realistic superconducting platforms, where crosstalk, calibration issues, and fluctuating unwanted couplings are unavoidable. It is important to note that the application of CDD generates a predictable time-independent $\mathrm{SU}\left(4\right)$ manifold on which we can then perform our geometric-control optimization strategies, without necessitating device-specific recalibration.

Once CDD stabilizes the drift Hamiltonian, in the subsequent optimal-control stage, the variational geodesic approach and the Monte Carlo method give coincident optimized trajectories, with essentially identical energetic costs and fidelities. The high agreement between these two independent approaches demonstrates that the sub-Riemannian geodesic solution is mathematically well-defined and numerically stable under gradient search. The fact that both methods produce the same geodesic trajectories, despite arbitrarily independent initial costates, shows the robustness of these optimization schemes converging to a unique optimal solution. The Krotov method proved successful in producing high-fidelity gates, but resulted in control functions with higher energetic cost.

Figures~\ref{CZ_graphs} and \ref{control_fields} show the strongly oscillatory structure of the Krotov solutions, which results in higher energetic costs. That the Krotov method will result in higher energy costs is not surprising, given that it does not optimize for energetic expenditure. The geometric variational and Monte Carlo methods, by construction, result in low energy costs, offering control trajectories naturally adapted to the geometry of the $\mathrm{SU}\left(4\right)$ group. However, the Krotov scheme produces more movement of the resulting trajectories through the manifold. The comparison among the three methodologies shows that the geodesic methods, designed to minimize energetic cost, also provide smoother control functions than the ones resulting from the standard Krotov method.

Of particular importance are the results summarized by Table~\ref{tab:tab1}. Even when one control direction is removed from the distribution, the procedure still finds solutions with unit fidelity, though at higher energetic costs. This is relevant since many superconducting platforms lack full three-axis control, instead using a combination of microwave drives and virtual rotations. The ability to maintain high fidelity under these typical constraints is a sign of the theoretical and experimental soundness of our methods, which are gifted with practical adaptability. Table~\ref{tab:tab1} illustrates how the energy cost increases depending on which axis is suppressed, revealing the algebraic redundancy of the controls in the $\mathrm{SU}\left(4\right)$ group. We see that suppression of the controls along a single direction is not prejudicial due to the bracket-generating ability of the operators along the other two directions to reinstate the action of the suppressed operator into the control Lie sub-algebra. The geometric variational, Monte Carlo, and Krotov algorithms, which comprise our methodology, produce trajectories that efficiently navigate the $\mathrm{SU}\left(4\right)$ manifold, even with reduced access to directional control.

Table~\ref{tab:tab1} also shows how the tests on the two additional entangling gates, the CX and a locally-equivalent arbitrary entangler, indicate that our approach is not specialized to the CZ gate, but applies to the complete set of $\mathrm{SU}\left(4\right)$ entangling gates. The geodesic equation is not partial to a specific target, except for its location in the manifold, resulting in reaching its optimized gates, irrespective of the local algebraic basis. This shows the universal applicability for generating two-qubit gates with high stability, despite noise and Hamiltonian imperfections. A closer look at the energetic cost and gate fidelity for the gate $U_\mathrm{R}$, Eq. (\ref{generic_U}), with each of the three components separately suppressed, is shown in detail in Fig. 4.

In summary, the results illustrate how our two-stage methodology operates. The application of CDD eliminates noise and drift systematically, leaving a stable Hamiltonian to be sculpted by the next stage. The optimal control methods, especially the Monte Carlo and Variational approaches, provide efficiency in generating minimal-energy and analytically structured control functions, leading to a desired target gate.

\section{Conclusions}

We have introduced a two-stage scheme for generating entangling gates with high fidelity and noise resilience. The method consists of applying the CDD method, followed by sub-Riemannian optimal control on $\mathrm{SU}\left(4\right).$ The CDD stage suppresses environmental couplings, crosstalk, and fluctuating residual terms in the native Hamiltonian, Eq.~(\ref{nativeH}). This renders the system stable and physically meaningful as a drift Hamiltonian. The next stage proceeds with geodesic optimization, learning the initial costate $\Lambda\left(0\right)$, containing the parameters for the optimal control trajectory.

Using a second variational principle, we derive a backward-propagated adjoint equation and an explicit gradient for $\Lambda\left(0\right)$, enabling the efficient and accurate minimization of gate infidelity. Three independent numerical methods confirm the effectiveness of this approach, with the variational and Monte Carlo geodesic methods consistently outperforming the well-established Krotov algorithm in energy cost. 

We illustrate the method on the CZ gate, CX gate, and a generic entangling gate, achieving virtually unit fidelity in all cases. Even restricting single-qubit control directions, in accordance with realistic platform limitations, the method is effective at generating perfect gates, with unsurprising increases in energetic costs. This performance confirms the generality of the geodesic control scheme and advances it as a powerful alternative to conventional pulse-design techniques.

The present study was based on an idealized $\mathrm{SU}\left(4\right)$ model without leakage. However, the geometric variational formalism naturally extends to larger models such as the $\mathrm{SU}\left(9\right)$ group, which shows its utility in the context of next-generation superconducting platforms, where leakage into higher transmon states is the dominant limitation. We anticipate that leakage suppression and higher-dimensional optimal control are straightforward and promising directions for future investigations.

In brief, the joint employment of CDD and geodesic variational optimal control constitutes a coherent, efficient, and experimentally realistic approach to design two-qubit gates, generating entangling gates with high fidelity, resilience to noise, and minimal energetic cost in a unified procedure.

% \textbf{\textcolor{red}{(Esta seção foi sugerida como template pelo
% ChatGPT4o)}}In this work, we introduced a two-stage control protocol
% for high-fidelity CZ gate implementation in a realistic two-qubit
% setup subject to environmental noise, spurious couplings, and crosstalk.

% The first stage applies a modified Continuous Dynamical Decoupling
% (CDD) scheme, which suppresses low-frequency noise and mitigates unwanted
% Hamiltonian components, transforming the native Hamiltonian into an
% effective drift Hamiltonian suitable for robust control.

% The second stage employs a geodesic-informed optimal control strategy
% based on the Pontryagin Maximum Principle, combined with three different
% numerical optimization approaches: Krotov's method, a Monte Carlo
% search, and a dual-PINN framework. This enables precise implementation
% of the CZ gate while minimizing control energy.

% Our results demonstrate that the proposed method achieves high gate
% fidelities even in the presence of realistic noise and parameter uncertainties,
% while offering a unified and systematic control framework. 

% Future extensions of this work include embedding the current SU(4)
% control model into an SU(9) framework to incorporate suppression of
% leakage to non-computational states, as well as generalization to
% multi-qubit architectures.

\appendix

\begin{acknowledgments}
A. H. da S. acknowledges financial support from Conselho Nacional de Desenvolvimento Cient\'ifico e Tecnol\'ogico (CNPq), project number 160849/2021-7. O.d.M. acknowledges financial support from Coordenação de Aperfeiçoamento de Pessoal de Nível Superior - Brasil (CAPES) – Finance Code  001.

R.d.J.N. and L.K.C. acknowledge support from Fundação de Amparo à Pesquisa do
Estado de São Paulo (FAPESP), projects number (2018/00796-3 and 2024/09298-7), and also
from the National Institute of Science and Technology for Quantum
Information (CNPq INCT-IQ 465469/2014-0) and the National Council
for Scientific and Technological Development (CNPq). L.K.C. also acknowledges the financial support of the National Institute of Science and Technology for Applied Quantum Computing through (CNPq 408884/2024-0). 
\end{acknowledgments}

\bibliography{refs.bib} 

\end{document}